\documentclass[twocolumn,prl,superscriptaddress,showpacs,byrevtex]{revtex4}
\usepackage{epsf,graphicx,amssymb,amsmath}
\usepackage[usenames]{color}

\begin{document}
\title{Observation of nonlinear dispersion relation and spatial\\ statistics of wave turbulence on the surface of a fluid}
\author{Eric Herbert}
\affiliation{Mati\`ere et Syst\`emes Complexes (MSC), Universit\'e Paris Diderot, CNRS (UMR 7057), 75 013 Paris, France}
\author{Nicolas Mordant}
\affiliation{Laboratoire de Physique Statistique, \'Ecole Normale Sup\'erieure, CNRS, 24, rue Lhomond, 75 005 Paris, France}
\author{Eric Falcon}
\affiliation{Mati\`ere et Syst\`emes Complexes (MSC), Universit\'e Paris Diderot, CNRS (UMR 7057), 75 013 Paris, France}

\date{\today}

\begin{abstract}
We report experiments on gravity-capillary wave turbulence on the surface of a fluid. The wave amplitudes are measured simultaneously in time and space using an optical method. The full space-time power spectrum shows that the wave energy is localized on several branches in the wave-vector-frequency space. The number of branches depend on the power injected within the waves. The measurement of the nonlinear dispersion relation is found to be well described by a law suggesting that the energy transfer mechanisms involved in wave turbulence are not only restricted to purely resonant interaction between nonlinear waves. The power-law scaling of the spatial spectrum and the probability distribution of the wave amplitudes at a given wave number are also measured and compared to the theoretical predictions. 
\end{abstract}
%47.35.-i Hydrodynamics waves
%05.45.-a  Nonlinear dynamics and chaos in statistical physics
%47.52.+j  Chaos in fluid dynamics\input{WTspatial2D.tex}
%47.27.-i	Turbulent flows 

\maketitle
Wave turbulence concerns the study of the statistical and dynamical properties of a set of numerous nonlinear interacting waves. It is an ubiquitous phenomenon observed in various situations from spin waves in solids, internal or surface waves in oceanography up to plasma waves in astrophysics (for recent reviews see~\cite{Falcon,Newell}). Wave turbulence theory, also called weak turbulence, predicts a wave energy cascade through the scales that can be derived analytically in nearly all fields of physics involving weakly nonlinear interacting waves in infinite systems \cite{ZakharovLivre}. However, few well-controled laboratory experiments have been performed so far, and show partial agreement with the theory~\cite{Falcon,Newell}. While most \emph{in situ} or laboratory measurements involve time signals at a fixed location, theoretical predictions often concern the Fourier space. An important challenge is thus to get a space-time measurement of the turbulent wave amplitudes (as recently achieved for elastic wave turbulence \cite{Nico09}), and thus to have a better understanding of the elementary dynamical processes involved in the energy cascade. Concerning wave turbulence on a fluid, previous results involve either 2D spatial measurements but not resolved in time (oceanography~ \cite{Hwang00} and laboratory experiments~~\cite{Jahne90}) or resolved in time but restricted to 1D space~\cite{Snouck09,Nazarenko10}. Here, we investigate 2D spatial and temporal statistics of wave turbulence on the surface of a fluid by using an optical profilometry technique. We perform a Fourier analysis of movies of the free-surface deformation and focus notably on the nonlinear dispersion relation. 
 
The experimental setup consists of a tank (46\, cm  $\times$ 36\,cm) filled with water (7\,cm deep). Surface waves are generated by the horizontal motion of two plunging rectangular wave makers (19\,cm in width and 2\,cm in depth). They are located at two corners of the same longest side of the tank, the vibration directions being perpendicular to each other \cite{Falcon07}. The wave makers are driven by two electromagnetic shakers submitted to a random forcing within a narrow low-frequency band (typically from 1 to 4 Hz). Typical maximal crest-to-trough wave amplitude ranges from 1 mm to 1.5 cm, and the wave mean steepness (ratio of crest-to-trough amplitude to its duration) ranges from 0.2 up to 3.3\,cm$/$s. This latter value corresponds to an injected power, $P$, 600 times greater than its value at the minimum forcing amplitude. This enables to access to linear, weakly and strongly nonlinear wave regimes. A Fourier transform profilometry method \cite{Takeda83,Cobelli09} provides the temporal evolution of the vertical deformation of the free-surface of the fluid over a significant spatial zone of the tank. Namely, a fringe pattern (wavelength $\lambda_f = 2.6$ or 5.2\,mm) is projected on the fluid surface by a video projector. When waves are generated, the vertical displacement of the free-surface leads to a phase shift of the pattern that is recorded by a camera. The deformation of the fluid surface $\eta(x,y,t)$  is then recovered by a 2D phase demodulation of each image of the recorded movie  \cite{Takeda83,Cobelli09}.  Movies are recorded with 1600 by 1200 pixels at $f_{acq}=50$ or 60 Hz during roughly 1 minute. The size of the recorded image is $25 \times 19$\,cm$^2$. To improve the contrast of the projected fringes on the fluid surface, a high concentrated white dye is added to the water bulk at an optimum concentration of 0.5\%\,v/v \cite{Cobelli09}. The surface tension of this dyed water is measured to be $\gamma =  32\pm 1$\,mN$/$m. Spatial and temporal resolutions of the measurement are 3$\lambda_f$ and $2/f_{acq}$, its linearity being insured for waves with sharp slopes up to 10 \cite{Cobelli09}. Possible underneath hydrodynamic turbulence generated in the bulk by the wave makers does not play a significant role on wave turbulence (similar results are found when the immersed length of the wave maker is changed).

\begin{figure}[t!]
\centerline{
\begin{tabular}{c}
  \epsfxsize=75mm
  \epsffile{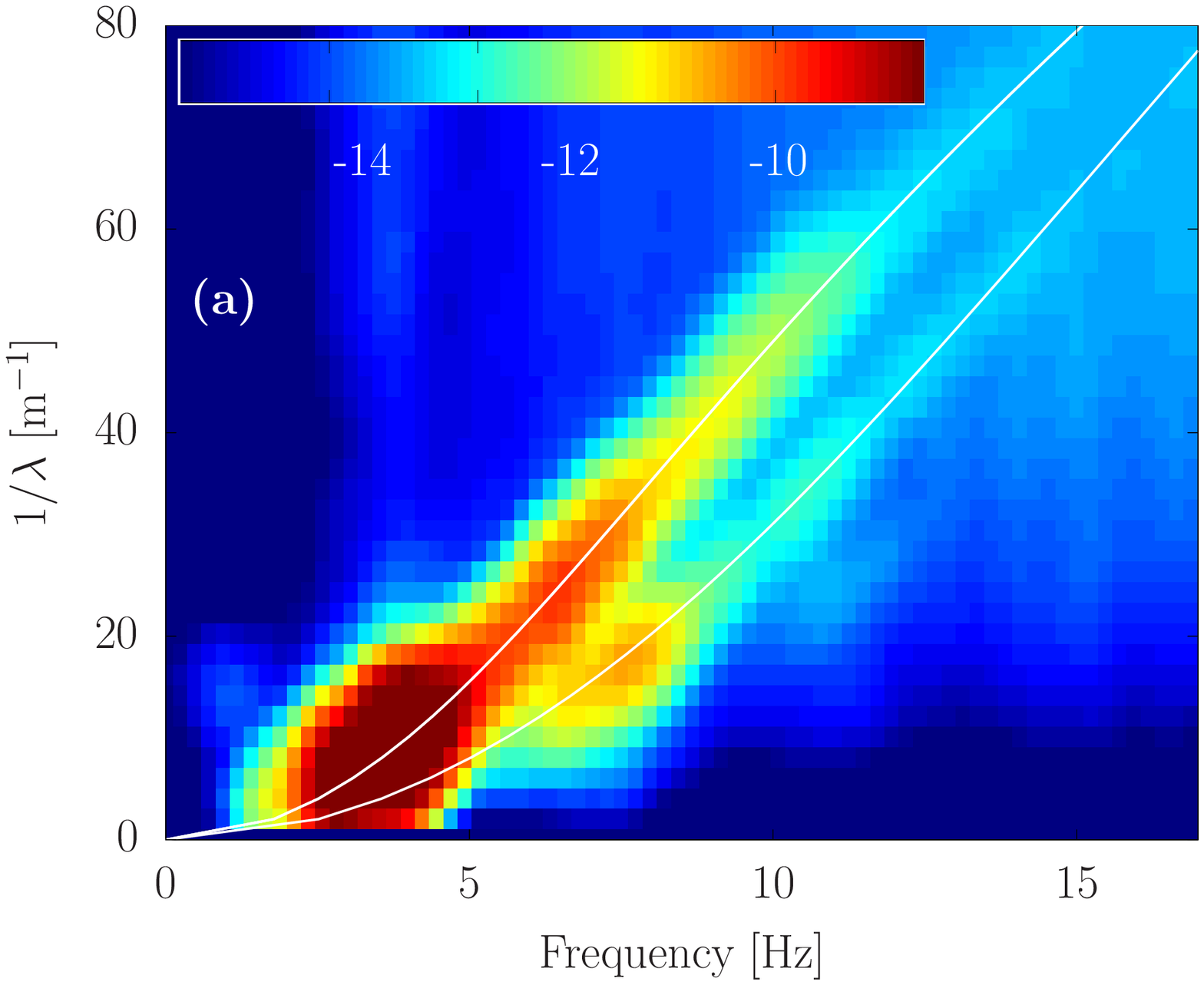}\\
  \epsfxsize=80mm
    \epsffile{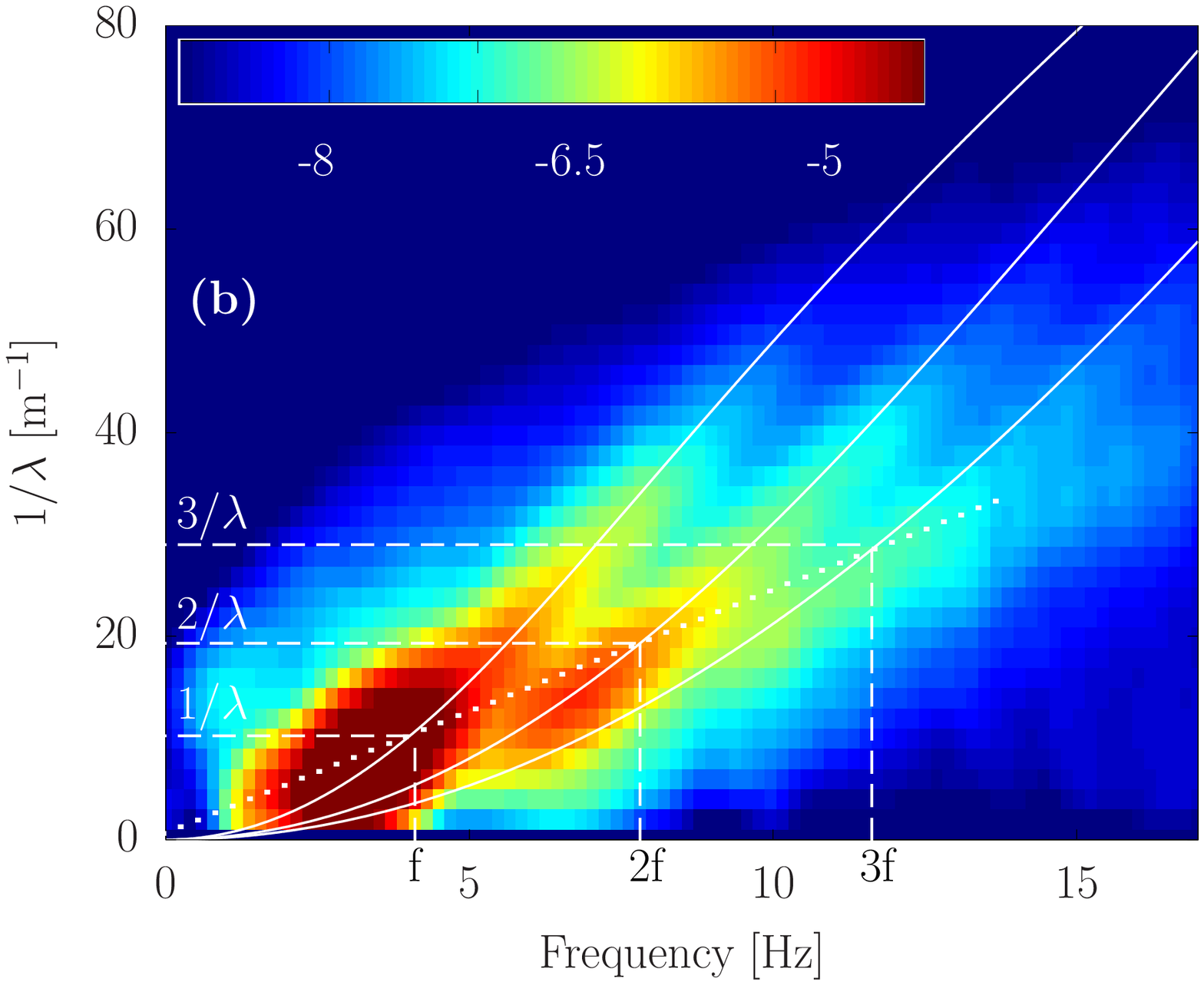}
\end{tabular}
}
\caption{(Color online) Space-time spectrum $E(k,f)$ of the vertical velocity of surface waves: moderate (a) and  strong (b) injected powers [$P^{1/2}$ = 5.2 (a)  and 24.7 (b) in arb. units]. Forcing: 1 - 4 Hz. Colors are log scaled. Solid white lines are $\Omega_N(k)$ with $N = 1$, 2, and 3 (see text). Slope of the dotted line corresponds to a constant phase velocity $\omega(k)/k$.}
\label{fig:spectres}
\end{figure}

The vertical velocity of the fluid surface $v(x,y,t)$ is obtained by differentiating the wave height movie in time. The full space-time power spectrum of the velocity $E(\mathbf{k} ,f)$ (a function of both the wave vector $\mathbf{k}$ and the frequency $f$) is then computed from multidimensional Fourier transform. By integrating $E(\mathbf{k} ,f)$ over all directions of $\mathbf{k}$, one obtains the velocity spectrum $E(k = ||\mathbf{k}||,f)$ displayed in Fig.~\ref{fig:spectres} for moderate and strong forcings. We observe that the energy injected at low frequencies cascades through the scales and is mainly localized on several branches in the ($\omega\equiv 2\pi f$, $k\equiv 2\pi/\lambda$) space. At low forcing amplitude (not shown here), only one branch occurs that corresponds to the linear gravity-capillary relation dispersion $\omega(k)=\sqrt{gk+(\gamma/\rho) k^3}$, with $g=9.81$ m/s$^{2}$ the acceleration of gravity, $\rho=1000$ kg/m$^3$ the fluid density. When the forcing is increased (see Fig.\ \ref{fig:spectres}a), a secondary branch appears below the linear dispersion relation (LDR). This branch is found to be well described by $\Omega_N(k)\equiv \sqrt{gNk + (\gamma/\rho) k^3/N}$ with $N=2$  with no adjustable parameter (see solid line). At higher forcing (see Fig.\ \ref{fig:spectres}b), a third branch appears following $\Omega_3(k)$. Thus, as the power injected in the wave system increases, the nonlinear wave interactions redistribute the wave energy on $N$ branches govern by  $\Omega_N(k)$, the nonlinear dispersion relation (NLDR). These secondary branches arise from the propagation of harmonics ($N\omega$,$Nk$) superimposed on a carrier long wave ($\omega$,$k$) and propagating with the phase velocity of the carrier (see below and Fig.\ \ref{fig:spectres}b). Note also that $\Omega_N(k)=N\omega_{k/N}$. Thus, at a fixed $k^{\star}$ corresponds $N$ peaks ($\omega_{k^{\star}}$, 2$\omega_{k^{\star}/2}$, 3$\omega_{k^{\star}/3}$, $\cdots$) in a frequency Fourier spectrum, i.e. a horizontal slice of Fig.\ \ref{fig:spectres}. This is consistent with  a two-peak frequency spectrum reported in a numerical simulation~\cite{Lvov06}. 

\begin{figure}[t!]
\centerline{
\epsfxsize=85mm
\epsffile{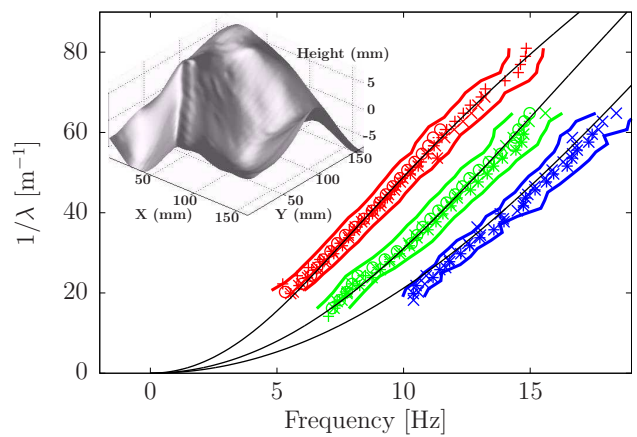}
}
\caption{(Color online) Nonlinear dispersion relation $k(f)$ computed from the lines of maximum energy of $E(k, f)$ for different forcings [$P^{1/2} = 1$ ($+$), $5.2$, ($\circ$), $10.5$ ($\times$), $24.7$ ($\ast$)]. Solid thick lines around each branch correspond to the branch width averaged for all forcings. Solid lines are $\Omega_N(k)$ with $N = 1$, 2, and 3 (see text - same as in Fig.~\ref{fig:spectres}). Inset: snapshot of the wave amplitudes at strong forcing ($P^{1/2} = 24.7$).}
\label{fig:RDs}
\end{figure}

At weak forcing, one observes linear gravity-capillary waves of gentle amplitudes that mix together. At strong forcing, steep long waves occur with sharp crest-ridges (see inset of Fig.\ \ref{fig:RDs}). Near the crests of these waves, high order harmonics are generated: small gravity-capillary waves superimposed on the long wave are observed (see also \cite{Falcon10}). These harmonics are called bound waves since they do not propagate with their own phase velocity but with the one of the carrier long wave \cite{LH63}, and thus leads to harmonics $\Omega_N(K)$ of velocity $\Omega_N(K)/K=\omega(k)/k$ where $K\equiv Nk$. They thus do not obey the linear dispersion relation which is consistent with the observation of secondary branches of the NLDR. This shows that other mechanisms than purely resonant wave interaction should be taken into account to describe the energy transfer across scales in wave turbulence.  

Let us now focus on the effect of the injected power $P$ on the location and the width of branches $\Omega_N(k)$. The maximum amplitude of each branch of the velocity spectrum $E(k,f)$ is extracted using the maximum of a Gaussian fit with respect to $k$ at a fixed $f$. For different $P$, the lines of maximum energy of each branch are shown in Fig.~\ref{fig:RDs} (symbols). Whatever the branch, the localized energy line is found to be independent of $P$: no measurable shift of these lines occurs in the ($k$,$\omega$) space. This differs from recent observation reported in elastic wave turbulence~\cite{Nico09} or in simulation~\cite{Lvov06}.  The widths of these branches are also plotted in Fig.~\ref{fig:RDs}. The width is defined by the rms value of the Gaussian fit. Whatever the branch, no significant evolution of the width is found when $P$ is increased. The width is also independent of the branch number within our experimental accuracy. The typical width [$\Delta (\lambda^{-1})$,$\Delta f$] centered on a point ($\lambda^{-1}$,$f$) of the NLDR is roughly (8 m$^{-1}$, 1.5 Hz), and could be ascribed to the typical nonlinear scale of wave mixing. However, one should be careful since $\Delta (\lambda^{-1})$ is close to the resolution of the discrete Fourier Transform ($\sim$ inverse of the image size $\sim 5$ m$^{-1}$). To sum up, the wave energy is redistributed on different branches of the NLDR of width that is independent of $P$ and of $N$. 

\begin{figure}[t!]
\centerline{
\epsfxsize=75mm 
\epsffile{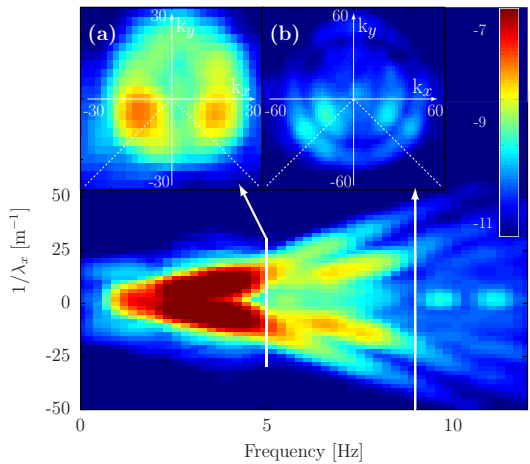}
}
\caption{(Color online) Space-time spectrum $E(k_x,f)$ of velocity at $P^{1/2}=10.5$ located at $k_y = 0$. Same forcing bandwidth as in Fig.\ \ref{fig:spectres}. Inset: space spectrum $E(k_x,k_y)$ located at $f =$5 Hz (a) and 9 Hz (b). Dashed lines: forcing directions. Log scaled colors are different for each plot. }
\label{fig:spectre_slice}
\end{figure}
\begin{figure}[t!]
\centerline{
\epsfxsize=80mm
\epsffile{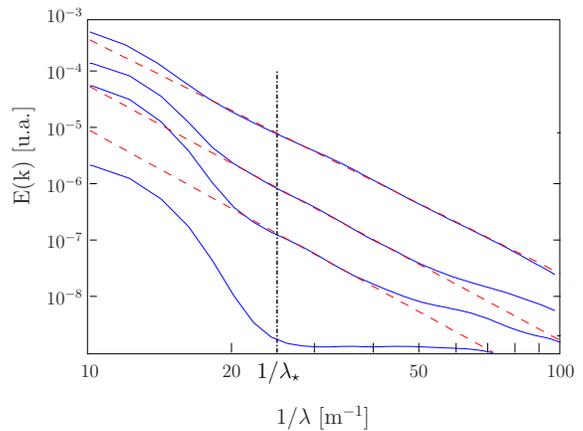}
}
\caption{(Color online) Spatial power spectrum $E(k)$ of the velocity for $P^{1/2} = $1, 5.2, 10.5, 24.7 (from bottom to top). Dashed lines have slopes $-4.6$, $-4.5$ and $-4.2$ (from bottom to top). Dot-dashed line: $\lambda^{-1}_\star= 25.2$\,m$^{-1}$ (see Fig. 5).}
\label{fig:kspectrum}
\end{figure}

Figure~\ref{fig:spectre_slice} shows different views of the full space-time Fourier spectrum of the velocity $E(\mathbf{k}, f)$. Main figure is $E(k_x, f)$, a slice at $k_y=0$ of the spectrum along the $x$-axis. Figures~\ref{fig:spectre_slice}a and \ref{fig:spectre_slice}b show $E(\mathbf{k})$ at two fixed frequencies (5 and 9\,Hz). In Fig.~\ref{fig:spectre_slice}a, the forcing reminisence appears as strong peaks in the direction of the wave makers (see dashed lines). Due to wave turbulence, energy cascades across scales as can be seen in the two continuous branches in the main Fig.~\ref{fig:spectre_slice}. Simultaneously, a spread of energy across angles is responsible for the continuous circles observed in the insets: the two concentric circles in Fig.~\ref{fig:spectre_slice}b correspond to two wave numbers given by both branches of the NLDR, whereas in Fig.~\ref{fig:spectre_slice}a one rather observes a single disc due to the overlapping of both branches of non-zero width (see above). Although not fully isotropic, this angular redistribution of energy is due to wave turbulence and is also visible as the symmetry between positive and negative $k_x$ in the main Fig.~\ref{fig:spectre_slice}.

The space spectrum $E(k\equiv ||\mathbf{k}||)$ of the velocity is then computed by summing the 3D space-time spectrum of $E(\mathbf{k},f)$  over all the directions of $\mathbf{k}$ and over $f$. Figure~\ref{fig:kspectrum} shows $E(k)$ when the forcing is increased. At high enough forcing, $E(k)$ is found to be scale invariant as expected for wave turbulence. The inertial range increases with the forcing, and $E(k) \sim k^{-n}$ with  $n \simeq 4.2$ over almost one decade in $k$ corresponding to $\lambda \sim$ few cm. Note that $n$ does not depend strongly on the forcing. Since one cannot compute $E(f)$ from $E(\mathbf{k},f)$ in a wide range of $f$ due to the strong steepness of the spectrum, one performs single point temporal measurements that shows a strong dependence of the power-law frequency exponent of the velocity spectrum on the forcing as already reported (typically from $f^{-5}$ to $f^{-2}$) \cite{Falcon07,Denissenko07}. These scalings suggest that the change of variable $k \leftrightarrow f$ using the LDR to estimate $E(k)$ from $E(f)$ is not valid in temporal measurements in hydrodynamics wave turbulence when strong nonlinear waves are involved. Indeed, this would lead to an estimated velocity spectrum from $k^{-3}$ to $k^{-3/2}$. Moreover, our velocity spectrum scaling, $E(k) \sim k^{-4.2}$, cannot be described by any of existing theories of wave turbulence taking into account either the presence of random-phased weakly nonlinear waves [$E_{theo}(k) \sim P^{1/2}k^{-1/2}$] \cite{Zakharov67Grav} or the dominance of coherent sharp wave crests [$E_{theo}(k) \sim k^{-1}$ to $k^{-3}$] \cite{Nazarenko10}. It is known numerically that the spatial spectrum exponent can change in case of anisotropy \cite{Polnikov01}. To confirm that our results do not depend on anisotropy, one computes $E(k)$ by summing $E(\mathbf{k},f)$ over different ($k_x$ , $k_y$) space regions: (i) over ($k_x$ , $k_y > 0$) where isotropy is observed, (ii) over ($k_x$ , $k_y < 0$) where anisotropy occurs due to the forcing (see insets of Fig.~\ \ref{fig:spectre_slice}), and (iii) all over ($k_x$ , $k_y$) space. The spatial spectrum $E(k) \sim k^{-z}$ computed over these different regions leads to $z=4.3$ (i), 3.8 (ii) and 4.2 (iii). Thus, the anisotropy does not play a significant role on the estimation of the spatial spectrum exponent. 

\begin{figure}[Ht!]
\centerline{
\epsfxsize=75mm
\epsffile{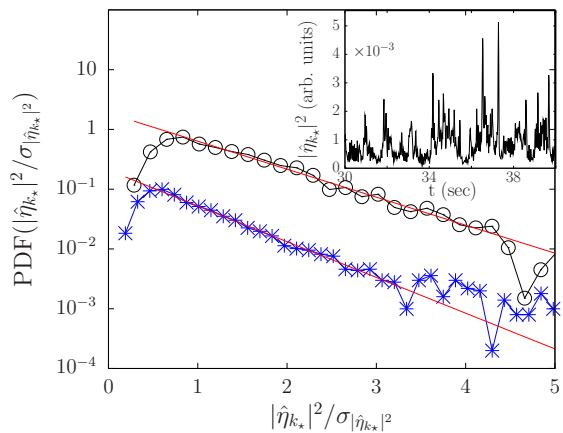}
}
\caption{(Color online) PDF of the Fourier wave amplitude, $|\hat{\eta}_{k_\star}|^2/\sigma_{|\hat{\eta}_{k_\star}|^2}$, at $k_\star\equiv 2\pi/\lambda_\star$ with $\lambda^{-1}_\star= 25.2 \pm 0.6$\,m$^{-1}$ (see Fig.~\ref{fig:kspectrum}) for two forcings  $P^{1/2} =5.2$ ($\circ$) and 10.5 ($\ast$).  $\langle |\hat{\eta}_{k_\star}|^2\rangle=1.1$ ($\circ$) and 1.3 ($\ast$). Solid lines have slopes $-0.46$ and $-0.58$. Curves have been shifted vertically for clarity. Inset: $|\hat{\eta}_{k_\star}|^2$ vs. time. $\lambda^{-1}_\star= 25.2 \pm 0.6 $\,m$^{-1}$. $P^{1/2} =24.7$.}
\label{fig:pdfspectreK}
\end{figure}

Finally, the probability density function (PDF) of the wave amplitude $\eta(x,y)$ is found to be roughly Gaussian whatever the forcing. One also computes the PDF of the Fourier amplitude $|\hat{\eta}_{k_\star}|^2$ of a wave component at a given wave number  $k_\star$. As shown in Fig.~\ref{fig:kspectrum}, we choose $k_\star \equiv 2\pi/\lambda_\star$ in the gravity regime  with  $\lambda_\star^{-1}=25.2 \pm 0.6$ m$^{-1}$ corresponding (using the LDR) to a frequency of $6.5$ Hz above the forcing ones.  For each image, the value of $|\hat{\eta}_{k_\star}|^2$ is extracted by averaging on 42 amplitudes found on a $\mathbf{k}$-space ring of radius $k_\star$. Iterating for all images leads to the temporal evolution of the Fourier amplitude of the mode $k_\star$ as shown in Fig.~\ref{fig:pdfspectreK}. This signal is strongly erratic and bursts of random large-amplitude occurs. Similar random bursts of Fourier amplitude has been reported in simulations \cite{Lvov06}, these bursts being correlated with phase jumps underlying strong nonlinear effects \cite{Lvov06}.  Although we are not able to measure the phase, this similarity is consistent with our above results underlying strong nonlinear effect. The PDF of the Fourier amplitude $|\hat{\eta}_{k_\star}|^2$, rescaled to its rms value $\sigma_{|\hat{\eta}_{k_\star}|^2}$, is then plotted in Fig.~\ref{fig:pdfspectreK} for two forcings. At low forcing, the PDF is roughly exponential as expected for random and uncorrelated waves. At higher forcing, the PDF remains Gaussian up to three standard deviations, whereas its tail shows a slight departure from this Gaussian.  Although more statistics are needed to characterize more deeply the PDF tail, this anomalously large probability of high Fourier mode amplitude is consistent with 1D spatial measurements \cite{Nazarenko10}, simulations \cite{Lvov06} and theory \cite{Choi05}. 

In conclusion, we have reported 2D spatial statistics of wave turbulence on the surface of a fluid. The power spectrum, the nonlinear dispersion relation and the PDF of the Fourier modes show strong effects of nonlinear waves involved in wave turbulence. This suggests that energy transfer mechanisms are not only restricted to resonant interactions between nonlinear waves, but also involve the formation of localized nonlinear structures (sharp-crested gravity waves) and of bound gravity-capillary waves. The wave spectrum scalings emphasize that the transition from $k$-space to $\omega$-space cannot be done according to the linear dispersion relation as usually performed in wave turbulence experiments.

\begin{acknowledgments}
This work has been supported by ANR Turbonde BLAN07-3-197846. We thank P. Chenevier.
\end{acknowledgments}
%%%%%%%%%%%%%%%%%%%%%%%%%%%%%%%%%%%%%%
%%%%%%%%%%%% REFERENCES %%%%%%%%%%%%%%%%%%
%%%%%%%%%%%%%%%%%%%%%%%%%%%%%%%%%%%%%%

\end{document}